\documentclass[a4paper,11pt]{article}
\pdfoutput=1 % if your are submitting a pdflatex (i.e. if you have
\usepackage{jcappub} % for details on the use of the package, please
\usepackage[T1]{fontenc} % if needed
\usepackage{booktabs}

\title{\boldmath Constraining light sterile neutrino mass with the BICEP2/Keck Array 2014 B-mode polarization data}

\author[a,b,1]{Shouvik Roy Choudhury,\note{Corresponding author.}}
\author[a,b]{Sandhya Choubey}

\affiliation[a]{Harish-Chandra Research Institute\\Chhatnag Road, Jhunsi, Allahabad 211019, India}
\affiliation[b]{Homi Bhabha National Institute\\ Training School Complex, Anushaktinagar, Mumbai - 400094, India}

\emailAdd{shouvikroychoudhury@hri.res.in}
\emailAdd{sandhya@hri.res.in}

\abstract{We explore the thermal light sterile neutrino situation from cosmological perspective in the $\Lambda \textrm{CDM} + r_{0.05} + N_{\textrm{eff}} + m^{\textrm{eff}}_{\textrm{s}}$ model using combinations of latest data sets available. Among CMB datasets, we use Planck 2015 temperature and low-$l$  polarization data and the latest data release on the B-mode polarization  from the BICEP2/Keck collaboration (BK14). We also use the latest Baryon Acoustic Oscillations (BAO) data from SDSS-III BOSS DR12, MGS, and 6dFS; and a Gaussian prior (HST) on the Hubble constant ($H_0 = 73.24 \pm 1.74$ km/sec/Mpc) from direct measurements by Hubble Space Telescope. We find that inclusion of BK14 data makes the constraints on the effective mass of sterile neutrino ($m^{\textrm{eff}}_{\textrm{s}}$) slightly stronger by preferring higher $\sigma_8$ values. The bound of $m^{\textrm{eff}}_{\textrm{s}} <$ 0.46 eV (95\% C.L.) is found for the combination of Planck 2015, BAO and BK14 datasets, whereas the bound is $m^{\textrm{eff}}_{\textrm{s}} <$ 0.53 eV (95\% C.L.) without the BK14 data. Our most aggressive bound of $m^{\textrm{eff}}_{\textrm{s}} <$ 0.28 eV (95\% C.L.) is obtained with Planck 2015, HST and BK14. Our analysis indicates that fully thermalized sterile neutrinos with mass $\sim 1$ eV are slightly more disfavoured with the inclusion of BK14 data. It also seems to make the agreement between Planck 2015 and CFHTLenS (weak gravitational lensing data) worse due to the higher $\sigma_8$ values.}

\begin{document}
\maketitle
\flushbottom

\section{Introduction}
\label{sec:1}
Sterile neutrinos still remain nothing short of an enigma in neutrino physics. Standard model predicts 3 massless neutrinos, while neutrino oscillation experiments have confirmed that neutrinos have mass, albeit very small. However, presence of anomalies in some short-baseline oscillation experiments \cite{Aguilar:2001ty,Aguilar-Arevalo:2013pmq,Mueller:2011nm,Mention:2011rk,Hayes:2013wra,Acero:2007su} have been explained with an extra species of neutrino, namely a sterile neutrino, of mass $\simeq$ 1 eV, which amply mixes with the active neutrinos but is uncharged under the standard model gauge group. Again, there are  analyses \cite{Kopp:2011qd,Kopp:2013vaa,Conrad:2012qt,Kristiansen:2013mza,Giunti:2013aea,Gariazzo:2017fdh,Capozzi:2016vac,Dentler:2017tkw,Dentler:2018sju} which indicate that all the results cannot be explained comfortably with the sterile neutrino hypothesis. A recent result \cite{Aguilar-Arevalo:2018gpe} from the MiniBooNE collaboration finds present electron neutrino and anti-neutrino appearance data still consistent with an extra sterile neutrino.

Apart from terrestrial oscillation experiments, in recent years, cosmology has become a very powerful probe of neutrino physics. In a situation where standard model of particle physics is augmented with only an extra sterile neutrino species, there are two parameters of utmost importance. One is the effective number of relativistic neutrino species, $N_{\textrm{eff}}$, whose theoretically predicted value, considering only the standard model of particle physics with 3 massless active neutrinos is $N_{\textrm{eff}}^{\textrm{SM}} = 3.046$ \cite{MANGANO2005221,PhysRevD.93.083522}, but is supposed to increase when contribution from the sterile neutrino is counted. The other is the effective mass of the sterile neutrino, $m^{\textrm{eff}}_{\textrm{s}} = \Delta N_{\textrm{eff}}^{3/4} m_{\textrm{s}}^{ph}$, where $\Delta N_{\textrm{eff}} = N_{\textrm{eff}} - 3.046$ and $m_{\textrm{s}}^{ph}$ is the physical mass of the sterile neutrino. Cosmology can provide strong constraints on these two parameters. $N_{\textrm{eff}}$, in general, can have contribution from any relativistic species which is not a photon, and hence it is not restricted only to the neutrino sector. Also, in certain scenarios like very low-reheating scenarios with sterile neutrinos \cite{Abazajian:2017tcc} or self-interacting sterile neutrinos \cite{Chu:2015ipa}, $\Delta N_{\textrm{eff}}$ can be negative. However, we do not consider such scenarios in this work, and consider only a non-interacting extra species of sterile neutrino. 

Provided we are only considering an extension to standard model with neutrino oscillations in a 3+1 scenario, as long as the sterile neutrino is of similar mass to an active neutrino and amply mixes with the active ones, its cosmological implications are identical to the active neutrino. Sufficient mixing will lead to almost complete thermalization \cite{Hannestad:2012ky,Melchiorri:2008gq}. However, even if there is partial thermalization, it will, in general, increase $N_{\textrm{eff}}$, leading to a delayed matter-radiation equality and a higher value of the Hubble parameter, $H(z_{dec})$, at the CMB decoupling (given other parameters are kept fixed). This has two main consequences \cite{Archidiacono:2013fha} on the CMB anisotropy power spectrum, first being an increase of the first peak of the spectrum due to early Integrated Sachs Wolfe (ISW) effect, and the second being a horizontal shift of the peaks towards higher multipoles. Along with a horizontal shift, there will also be a vertical shift which will decrease the amplitude of the peaks at high multipoles, a phenomenon related to Silk damping. These effects of an additional relativistic sterile neutrino can be partially compensated if other cosmological parameters are simultaneously varied. For example, if the total matter density $\omega_{m}$ is also increased without altering the baryon density, the redshift of matter-radiation equality can be kept fixed. These degeneracies tend to degrade the constraints on $N_{\textrm{eff}}$. However, the CMB power spectra won't be exactly the same even after such adjustments with other parameters, especially because of the neutrino anisotropic stress arising from the quadrupole moment of the cosmic neutrino background temperature anisotropies which alters the gravitational potentials \cite{Bashinsky:2003tk,Hannestad:2004qu}. Hence constraints can be put on $N_{\textrm{eff}}$ from CMB power spectra data.

If a light sterile neutrino has a mass $\simeq$ 1 eV, it only starts to become non-relativistic during CMB, and hence the effect of the mass is not strong on CMB power spectra. Sterile neutrinos with masses much smaller than 1 eV will have negligible effect on CMB power sprectra. However, when CMB power spectra data is used with other cosmological observations like constraining the Hubble parameter from direct measurements via a Gaussian prior or using the Baryon Acoustic Oscillation (BAO) data or both, better bounds on the mass of the sterile neutrino can be obtained \cite{Lesgourgues:2014zoa}. A plethora of papers \cite{Lesgourgues:2014zoa,LESGOURGUES2006307,doi:10.1146/annurev-nucl-102010-130252,Lesgourgues-Pastor,ABAZAJIAN201566,1475-7516-2017-02-052,Gariazzo:2015rra} are available on the effects of neutrino masses on cosmology. Current bounds on sterile neutrinos from cosmological data imply that fully thermalized sterile neutrinos of mass $\simeq$ 1 eV are disfavoured and can only be accommodated with partial thermalization. See previous analyses on constraining sterile neutrino properties with cosmological data \cite{Hamann:2011ge,Ko:2014bka,Archidiacono:2014apa,Zhang:2014nta,Archidiacono:2014nda,Li:2014dja,Zhang:2014ifa,Zhang:2014lfa,Li:2015poa,Archidiacono:2012ri,Archidiacono:2013xxa,Gariazzo:2013gua,Dvorkin:2014lea,Giusarma:2014zza,Feng:2017nss,Feng:2017mfs,Feng:2017usu,Song:2018zyl,Heavens:2017hkr,Forastieri:2017oma,Zhao:2017urm,Bridle:2016isd,Kumar:2017dnp}.

In this paper, we have, for the first time, used the BK14 data, the data on the B-mode polarization of CMB from BICEP2/Keck collaboration, to constrain the parameters associated with sterile neutrinos in an extended $\Lambda \textrm{CDM}$ model, which can be simply denoted with $\Lambda \textrm{CDM} + r_{0.05} + N_{\textrm{eff}} + m^{\textrm{eff}}_{\textrm{s}}$. 
BK14 constrains the tensor-to-scalar ratio to $r_{0.05}<$ 0.07 at 95\% C.L, when combined with Planck 2015 and other datasets \cite{PhysRevLett.116.031302}; while exclusion of the BK14 data leads to a significantly less strong bound of $r_{0.05}<$ 0.12 \cite{Planck2015}. BK14 data also contains information on gravitational lensing. Thus we expect this data to affect the constraints on the sterile neutrino parameters. We also provide results with $N_{\textrm{eff}}$ fixed at 4.046 and 3.5 separately, i.e., assuming full and partial thermalization of the sterile neutrinos respectively, and this model is denoted as $\Lambda \textrm{CDM} + r_{0.05} + m^{\textrm{eff}}_{\textrm{s}}$.

This paper has the following structure: in Section~\ref{sec:2} we provide details about our model parameters and other analysis details and briefly describe the datasets used, in Section~\ref{sec:3} we provide the results of our analysis, and we conclude in Section~\ref{sec:4}.

\section{Cosmological Analysis}
\label{sec:2}

\subsection{Model}
\label{sec:2:1}

Below we list the vector of parameters we have varied in this work in two cosmological models.

For $\Lambda \textrm{CDM} + r_{0.05} + N_{\textrm{eff}} + m^{\textrm{eff}}_{\textrm{s}}$ model: 
\begin{equation}\label{eqn:1}
\theta \equiv \left[\omega_c, ~\omega_b, ~\Theta_s,~\tau, ~n_s, ~ln [ 10^{10} A_s], r_{0.05}, N_{\textrm{eff}}, m^{\textrm{eff}}_{\textrm{s}} \right].
\end{equation} 

For $\Lambda \textrm{CDM} + r_{0.05} + m^{\textrm{eff}}_{\textrm{s}}$ model:
\begin{equation}\label{eqn:2}
\theta \equiv \left[\omega_c, ~\omega_b, ~\Theta_s,~\tau, ~n_s, ~ln [ 10^{10} A_s], r_{0.05}, m^{\textrm{eff}}_{\textrm{s}} \right].
\end{equation} 
with $N_{\textrm{eff}}$ fixed to the value 4.046, which corresponds to full thermalization of the sterile neutrino with active neutrinos and to the value 3.5, which corresponds to partial thermalization. 

The first six parameters correspond to the $\Lambda \textrm{CDM}$ model. Here $\omega_c = \Omega_c h^2$  and  $\omega_b = \Omega_b h^2$ are the physical cold dark matter and baryon densities at present, respectively. $\Theta_s$ is the angular sound horizon, i.e., the ratio between sound horizon and the angular diameter distance at decoupling. $\tau$ is the reionization optical depth. $n_s$ and $A_s$ are the power-law spectral index and power of the inflationary power spectrum, respectively, at the pivot scale of $k_* = 0.05 h$ Mpc$^{-1}$. 

$r_{0.05}$ is the tensor-to-scalar ratio, also defined at the pivot scale of $k_*=0.05h$  Mpc$^{-1}$. $N_{\textrm{eff}}$, effective number of relativistic species which are not photons, is given by,
\begin{equation}
\rho_r = \frac{\pi^2}{15}\left[ 1+ \frac{7}{8} \left(\frac{4}{11}\right)^{\frac{4}{3}} N_{\textrm{eff}} \right] T_{\gamma}^4,
\end{equation}
where $T_{\gamma}$ is the temperature of the photons and $\rho_r$ is the radiation density. In our model, apart from photons, we only have 3 active and one sterile neutrino as relativistic species. The sterile neutrino is assumed not to have any self-interactions, or interactions with other particle species. In our work, we have fixed the active neutrino sector to give a contribution of $N_{\textrm{eff}}^{\textrm{SM}} =$3.046 to $N_{\textrm{eff}}$, with two massless and one massive neutrino with mass of 0.06 eV. Thus the contribution to $N_{\textrm{eff}}$ from the sterile species is simply $\Delta N_{\textrm{eff}} = N_{\textrm{eff}} - 3.046$. Here it should be mentioned that in certain scenarios like low reheating  Note that while we use the value 3.046 which is predominant in literature, a recent study \cite{deSalas:2016ztq}  had found $N_{\textrm{eff}}^{\textrm{SM}} =$3.045.

When the sterile neutrino is relativistic at early times, assuming the only radiation species are photons and neutrinos, contribution of a light sterile neutrino to $N_{\textrm{eff}}$ is given by \cite{Acero:2008rh},
\begin{equation}
\Delta N_{\textrm{eff}} =\left[\frac{7}{8}\frac{\pi^2}{15}T_{\nu}^{4} \right]^{-1} \frac{1}{\pi^2} \int dp ~p^3 ~f_s (p),
\end{equation}
where $T_\nu$ is active neutrino temperature, $p$ is the neutrino momentum, and $f_s(p)$ is momentum distribution function of the sterile neutrino. At late times its energy density is parametrized as an effective mass \cite{Acero:2008rh,Ade:2013zuv}:
\begin{equation}
\omega_s \equiv \Omega_s h^2 = \frac{m_s^{\textrm{eff}}}{94.1 \textrm{eV}} = \frac{h^2 m_s^{\textrm{ph}}}{\pi^2 \rho_{c}} \int dp~ p^2 ~f_s (p),
\end{equation}
where $\rho_{c}$ is the critical density, $\Omega_s h^2$ is the sterile neutrino energy density. Since sterile neutrinos don't have electroweak interactions and they have mixing with the active neutrinos, they cannot decouple after the decoupling of active neutrinos. Active neutrinos decouple at a temperature $T \sim 1$ MeV, when all of them are relativistic. Hence $f_s (p)$ doesn't depend on the physical mass of the sterile neutrino, $m_s^{\textrm{ph}}$. However $f_s (p)$ depends on the production mechanism of the light sterile neutrino. If the production is through a thermal process, one can simply write $f_s (p) = (e^{p/T_s}+1)^{-1}$, the usual Fermi-Dirac distribution function, where $T_s$ is the sterile neutrino temperature. In this case, it can be shown that,
\begin{equation}
\label{eqn:3}
m^{\textrm{eff}}_{\textrm{s}} = \Delta N_{\textrm{eff}}^{3/4} m_{\textrm{s}}^{ph};~~~~~~~~\Delta N_{\textrm{eff}} = \left(\frac{T_s}{T_{\nu}}\right)^4.
\end{equation}
Non-thermal production, on the other hand, can lead to various possible scenarios. One of the popular scenarios is the Dodelson-Widrow (DW) mechanism \cite{Dodelson:1993je}, for which $f_s (p) = \beta (e^{p/T_{\nu}}+1)^{-1}$, where $\beta$ is a normalization factor. In this case, one gets \cite{Acero:2008rh},

\begin{equation}
m^{\textrm{eff}}_{\textrm{s}} = \Delta N_{\textrm{eff}}~ m_{\textrm{s}}^{ph};~~~~~~~~\Delta N_{\textrm{eff}} =\beta.
\end{equation}

So, the $m^{\textrm{eff}}_{\textrm{s}}$ parametrization can accommodate two different scenarios of sterile neutrino production. Also notice that in the $\Lambda \textrm{CDM} + r_{0.05} + m^{\textrm{eff}}_{\textrm{s}}$ model, fixing $N_{\textrm{eff}} =$ 4.046 leads to $m^{\textrm{eff}}_{\textrm{s}}$ being same as $m_{\textrm{s}}^{ph}$.

In our work, we conduct a Bayesian analysis to derive constraints on the sterile neutrino parameters.  For all the parameters listed in Eq.~(\ref{eqn:1}), and Eq.~(\ref{eqn:2}), we impose flat priors. We also limit the physical mass of the sterile neutrino to $m_{\textrm{s}}^{ph} \leq$ 10 eV. The prior ranges are provided on the  Table~\ref{table:1}. We run chains using the November 2016 version of the Markov Chain Monte Carlo (MCMC) sampler CosmoMC \cite{PhysRevD.66.103511} which incorporates CAMB \cite{Lewis:1999bs} as the Boltzmann code and the Gelman and Rubin statistics \cite{doi:10.1080/10618600.1998.10474787} to estimate the convergence of chains.
\begin{table}[tbp]
\centering
\begin{tabular}{|l|c|} 
\hline
%\toprule
%\toprule
Parameter & Prior\\
\hline
%\vspace{0.2cm} 
$\omega_c$ & [0.001,0.99] \\
%\vspace{0.2cm}
%\hline 
$\omega_b$ & [0.005,0.1]  \\
%\vspace{0.2cm}
%\hline
$\Theta_s$ & [0.5,10] \\
%\vspace{0.2cm}
%\hline
$\tau$ & [0.01,0.8]\\
%\vspace{0.2cm}
%\hline
$n_s$ & [0.8,1.2] \\
%\vspace{0.2cm}
%\hline
ln $[10^{10} A_s]$ & [2,4] \\
%\vspace{0.2cm}
%\hline
$r_{0.05}$ & [0,2] \\
%\vspace{0.2cm}
%\hline
$N_{\textrm{eff}}$ & [3.046,7] \\
%\vspace{0.2cm}
$m^{\textrm{eff}}_{\textrm{s}}$ & [0,3]\\
%\bottomrule
%\bottomrule
\hline

\end{tabular}
\caption{\label{table:1} Flat priors on cosmological parameters included in this work.}
\end{table}

\subsection{Datasets}
\label{sec:2:2}

We use separate combinations of the following datasets:

\emph{Cosmic Microwave Background: Planck 2015}:

Measurements of the CMB temperature and low-$l$ polarization from Planck 2015 \cite{Planck2015-I} are used. We consider the high-$l$ (30 $\leq$ $l$ $\leq$ 2508) TT likelihood, and also the low-$l$ (2 $\leq$ $l$ $\leq$ 29) TT likelihood. We refer to this combination as TT. We also include the Planck polarization data in the low-$l$ (2 $\leq$ $l$ $\leq$ 29) likelihood, and denote this as lowP. We also use the Planck lensing potential measurements via reconstruction through the four-point correlation functions of the Planck CMB data \cite{Ade:2015zua}. We call this simply as lensing. Residual systematics may be present in the the Planck 2015 high-$l$ polarization data \cite{Planck2015}, so we refrain from using it.\\

\emph{B Mode Polarization data of CMB}: 

Considering the B-mode polarization of CMB, we incorporate the recent dataset publicly available from BICEP2/Keck collaboration which includes all data (multipole range: $20 < l < 330$) taken up to and including 2014 \cite{PhysRevLett.116.031302}. This dataset is referred to as BK14. \\

\emph{Baryon Acoustic Oscillations (BAO) Measurements and Related Galaxy Cluster data}:

In this analysis, we include measurements of the BAO signal obtained from different galaxy surveys. We make use of the SDSS-III BOSS DR12 \cite{doi:10.1093/mnras/stx721} LOWZ and CMASS galaxy samples at $z_{\textrm{eff}} =$ 0.38, 0.51 and 0.61, the DR7 Main Galaxy Sample (MGS) at $z_{\textrm{eff}} = 0.15$ \cite{doi:10.1093/mnras/stv154}, and the 6dFGS survey at $z_{\textrm{eff}} = 0.106$ \cite{doi:10.1111/j.1365-2966.2011.19250.x}. We call this complete combination as BAO. Here $z_{\textrm{eff}}$ is the effective redshift of a survey. \\

\emph{Hubble Parameter Measurements}: 

We use a Gaussian prior of $73.24 \pm 1.74$ km/sec/Mpc  on $H_0$, which is a recent 2.4\% determination of the local value of the Hubble parameter by \cite{0004-637X-826-1-56} which combines the anchor NGC 4258, Milky Way and LMC Cepheids. We denote this prior as HST.\\

\section{Results}
\label{sec:3}
For convenience, we have separated the results in two subsections for the the two different models. The description of models and datasets are given at section \ref{sec:2:1} and section \ref{sec:2:2}, respectively. We have presented the results, first in the $\Lambda \textrm{CDM} + r_{0.05} + N_{\textrm{eff}} + m^{\textrm{eff}}_{\textrm{s}}$ model, and then in the $\Lambda \textrm{CDM} + r_{0.05} + m^{\textrm{eff}}_{\textrm{s}}$ model. All the marginalized limits quoted in the text or tables are at 68\% C.L. whereas upper limits are quoted at 95\% C.L., unless otherwise specified.

\subsection{Results for $\Lambda \textrm{CDM} + r_{0.05} + N_{\textrm{eff}} + m^{\textrm{eff}}_{\textrm{s}}$ model}
\label{sec:3:1}

\begin{table}
	
	\centering
%	\small\addtolength{\tabcolsep}{-4pt}
    \resizebox{\textwidth}{!}{
	\begin{tabular}{cccccc}
		\toprule
		\toprule
		 \vspace{0.2cm}
		Parameter &    TT+lowP  &    TT+lowP  &    TT+lowP &    TT+lowP     &  TT+lowP\\ \vspace{0.2cm}
		          &             &      +BAO   &      +HST  &    +HST+BAO    &  +HST+BAO+lensing\\
		\midrule
		\hspace{1mm}
		\vspace{ 0.2cm}
		$m^{\textrm{eff}}_{\textrm{s}}$ (eV)  &    $<0.78$  &    $<0.53$  &    $<0.34$  &    $<0.36$  &    $<0.40$  \\ \vspace{ 0.2cm}
		
	 	$N_{\textrm{eff}}$  &    $<3.78$  &    $<3.75$  &    $3.63\pm0.21$  &    $3.59\pm0.22$ &   $3.60^{+0.21}_{-0.24}$  \\ \vspace{ 0.2cm}
	 	$r_{0.05}$   &    $<0.127$  &    $<0.129$  &    $<0.151$  &    $<0.148$ &    $<0.155$  \\ \vspace{ 0.2cm}	
	
	 	$H_0$ (km/sec/Mpc)  &    $68.35^{+1.23}_{-2.50}$  &    $69.14^{+0.89}_{-1.59}$  &    $71.77^{+1.63}_{-1.64}$  &     $70.79_{-1.20}^{+1.19}$  &     $70.78\pm1.21$   \\ \vspace{ 0.2cm}
		
	 	$\sigma_8$  &    $0.802^{+0.040}_{-0.029}$  &    $0.815^{+0.029}_{-0.023}$  &    $0.836^{+0.029}_{-0.021}$  &    $0.828^{+0.029}_{-0.023}$ &    $0.816^{+0.020}_{-0.016}$    \\ 
		\bottomrule
		\bottomrule
	\end{tabular}}
	\caption{\label{table:2} \footnotesize Bounds on cosmological parameters in the $\Lambda \textrm{CDM} + r_{0.05} + N_{\textrm{eff}} + m^{\textrm{eff}}_{\textrm{s}}$ model without BK14 data. Marginalized limits are given at 68\% C.L. whereas upper limits are given at 95\% C.L.. Note that $H_0$ and $\sigma_8$ are derived parameters.}
\end{table}	

\begin{table}
	
	\centering
%	\small\addtolength{\tabcolsep}{-4pt}
    \resizebox{\textwidth}{!}{
	\begin{tabular}{cccccc}
		\toprule
		\toprule
		 \vspace{0.2cm}
		Parameter &    TT+lowP+BK14  &    TT+lowP+BK14  &    TT+lowP+BK14 &    TT+lowP+BK14     &  TT+lowP+BK14\\ \vspace{0.2cm}
		          &             &      +BAO   &      +HST  &    +HST+BAO    &  +HST+BAO+lensing\\
		\midrule
		\hspace{1mm}
		\vspace{ 0.2cm}
		$m^{\textrm{eff}}_{\textrm{s}}$ (eV)  &    $<0.68$  &    $<0.46$  &    $<0.28$  &    $<0.30$  &    $<0.35$  \\ \vspace{ 0.2cm}
		
	 	$N_{\textrm{eff}}$  &    $<3.76$  &    $<3.74$  &    $3.63\pm0.21$  &    $3.59\pm0.21$ &   $3.59^{+0.21}_{-0.23}$  \\ \vspace{ 0.2cm}
	 	$r_{0.05}$   &    $<0.068$  &    $<0.070$  &    $<0.073$  &    $<0.072$ &    $<0.078$  \\ \vspace{ 0.2cm}	
	
	 	$H_0$ (km/sec/Mpc)  &    $68.31^{+1.25}_{-2.48}$  &    $69.16^{+0.95}_{-1.61}$  &    $71.73\pm1.62$  &     $70.84\pm1.20$  &     $70.75^{+1.17}_{-1.18}$   \\ \vspace{ 0.2cm}
		
	 	$\sigma_8$  &    $0.814^{+0.036}_{-0.027}$  &    $0.825^{+0.027}_{-0.021}$  &    $0.846^{+0.026}_{-0.020}$  &    $0.841^{+0.025}_{-0.021}$ &    $0.820^{+0.019}_{-0.015}$    \\ 
		\bottomrule
		\bottomrule
	\end{tabular}}
	\caption{\label{table:3} \footnotesize Bounds on cosmological parameters in the $\Lambda \textrm{CDM} + r_{0.05} + N_{\textrm{eff}} + m^{\textrm{eff}}_{\textrm{s}}$ model with BK14 data. Marginalized limits are given at 68\% C.L. whereas upper limits are given at 95\% C.L.. Note that $H_0$ and $\sigma_8$ are derived parameters.}
\end{table}	

In this section, we present the results for the $\Lambda \textrm{CDM} + r_{0.05} + N_{\textrm{eff}} + m^{\textrm{eff}}_{\textrm{s}}$ model. In Table~\ref{table:2}  we have provided results without BK14 data, whereas, in Table~\ref{table:3}, the results are with BK14, to compare. We have presented constraints on the three parameters $r_{0.05}$, $N_{\textrm{eff}}$, and $m^{\textrm{eff}}_{\textrm{s}}$. with which we have extended the $\Lambda \textrm{CDM}$ model, and also two derived parameters $H_0$ and $\sigma_8$, which are important in constraining the sterile neutrino mass.   

\begin{figure}[tbp]
\centering % \begin{center}/\end{center} takes some additional vertical space
\includegraphics[width=.4963\linewidth]{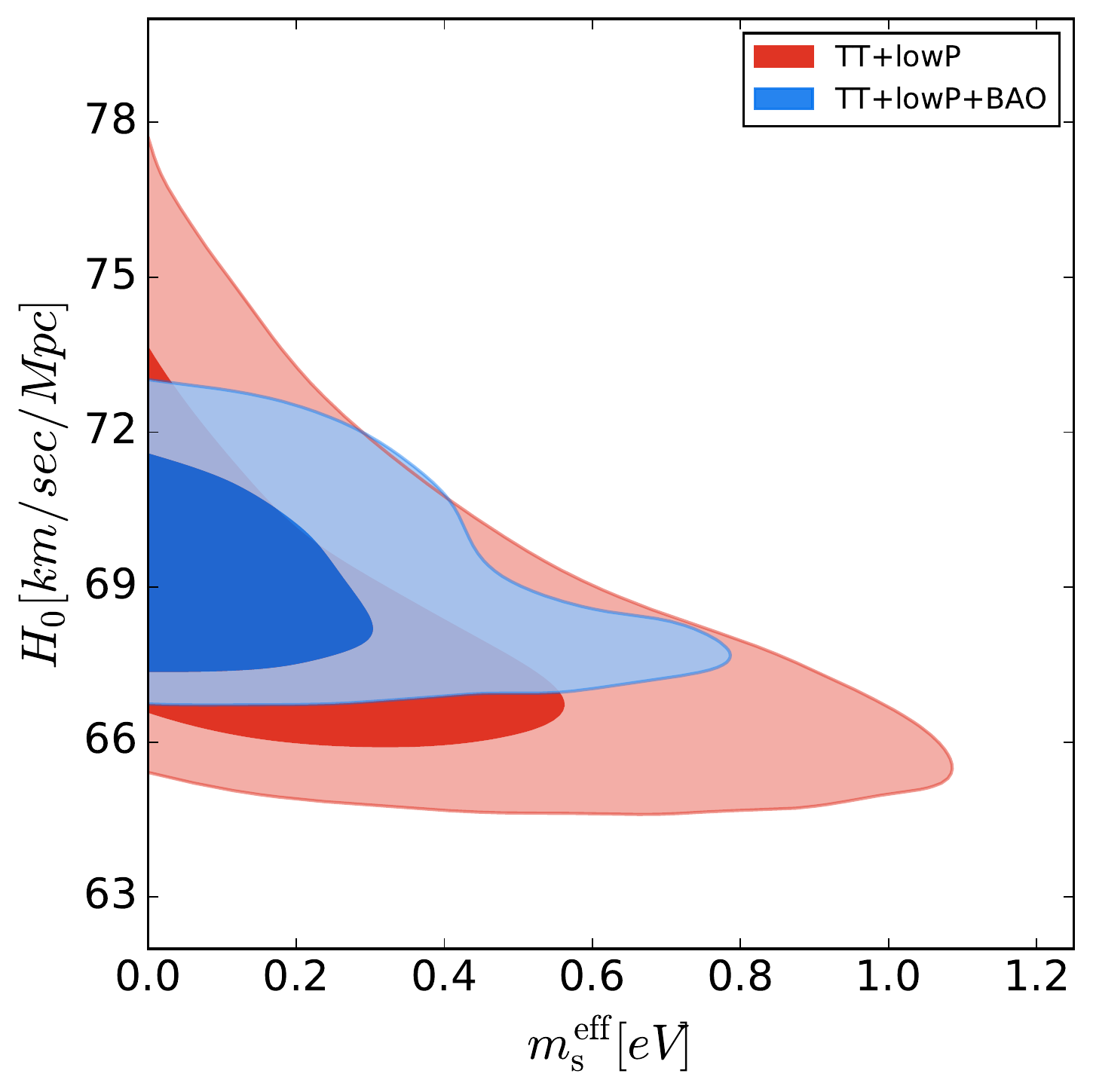}
\hfill
\includegraphics[width=.4963\linewidth]{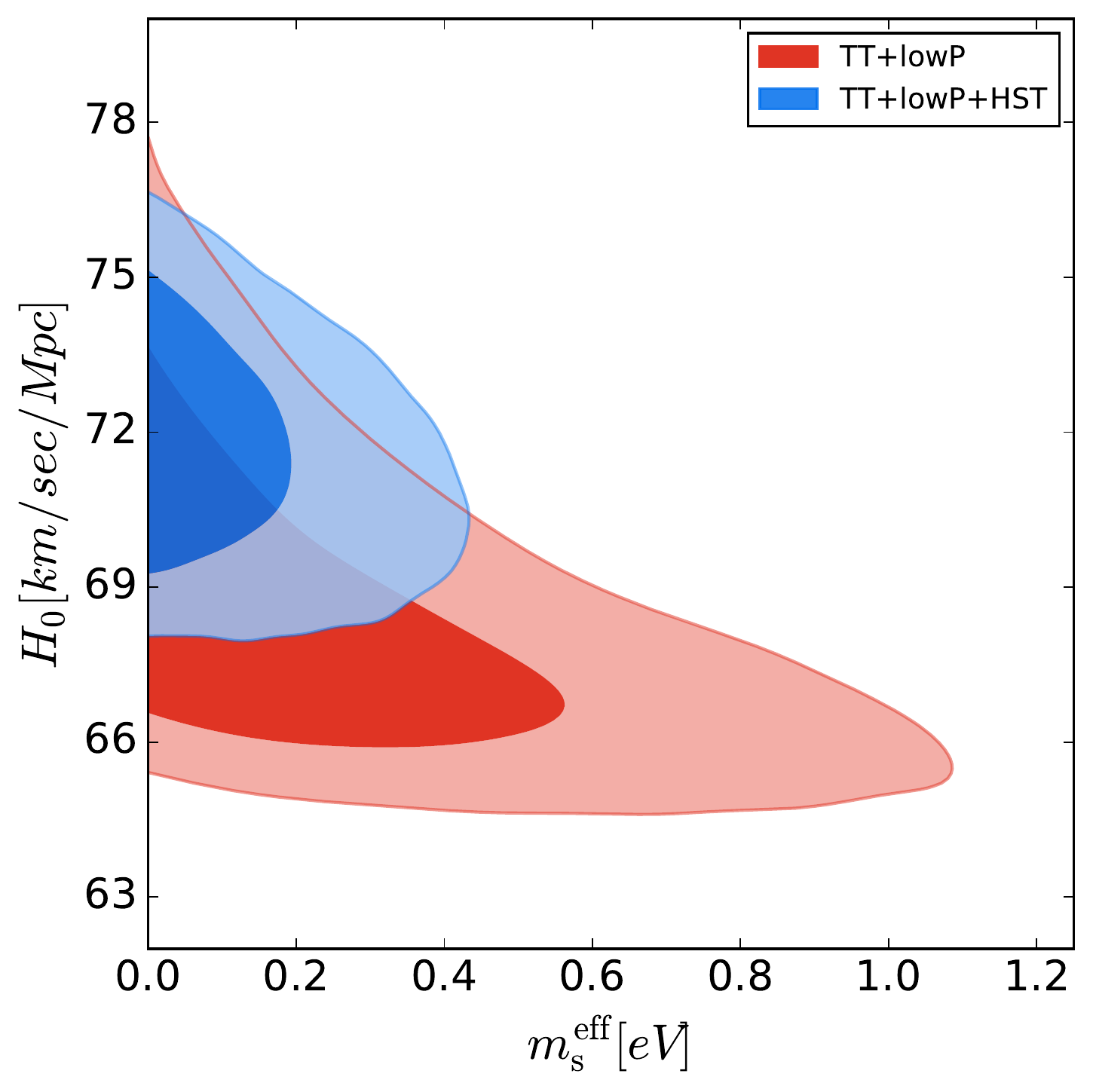}
% "\includegraphics" is very powerful; the graphicx package is already loaded
\caption{\label{fig:1}1$\sigma$ and 2$\sigma$ marginalized  contours for $H_0$ [km/sec/Mpc] vs. $m^{\textrm{eff}}_{\textrm{s}}$ [eV] in the $\Lambda \textrm{CDM} + r_{0.05} + N_{\textrm{eff}} + m^{\textrm{eff}}_{\textrm{s}}$ model with the following combinations: TT+lowP, TT+lowP+BAO, and TT+lowP+HST. Both BAO and HST data decrease the correlation between the two parameters significantly.}
\end{figure}

With only TT+lowP, we see that the bound on the sterile mass is relaxed at $m^{\textrm{eff}}_{\textrm{s}}<$ 0.78 eV. The bound gets tightened with BAO data, which partially breaks the degeneracy between $m^{\textrm{eff}}_{\textrm{s}}$ and $H_0$ present in the TT+lowP data, by rejecting lower values of $H_0$ \cite{Choudhury:2018byy,Hou:2012xq} and leads to a bound of $m^{\textrm{eff}}_{\textrm{s}}<$ 0.53 eV. This effect can be seen pictorially in Figure~\ref{fig:1} where addition of BAO data leads to a significantly smaller magnitude of anti-correlation between $m^{\textrm{eff}}_{\textrm{s}}$ and $H_0$. The HST prior also breaks the degeneracy partially, as can be seen in Figure~\ref{fig:1}. However, the $H_0$ values preferred by the HST prior are larger than BAO, which leads to a preference to even smaller masses ($m^{\textrm{eff}}_{\textrm{s}}<$ 0.34 eV) to keep the comoving distance to the surface of last scattering fixed \cite{Choudhury:2018byy}. Adding HST and BAO together with CMB however does not provide better bound than CMB+HST. Also, the lensing data degrades the bound on $m^{\textrm{eff}}_{\textrm{s}}$. We note that CMB and/or BAO data do not allow full thermalization of sterile neutrinos. However, at 95\% C.L., with TT+lowP+HST, we obtained a $N_{\textrm{eff}} = 3.63^{+0.44}_{-0.42}$. Such high values of $N_{\textrm{eff}}$ disallow the standard model prediction of $N_{\textrm{eff}}^{SM} = 3.046$ at 95\% C.L. but allow $N_{\textrm{eff}} = 4.046$, i.e., full thermalization. 
On the other hand, it is also imperative to consider recent constraints on $N_{\textrm{eff}}$ coming from Big Bang Nucleosynthesis (BBN). Planck 2018 results \cite{Aghanim:2018eyx} have provided bound of $N_{\textrm{eff}} = 2.95^{+0.56}_{-0.52}$ (95\% C.L.) (which is independent of the details of the CMB spectra at high multipoles) by combining the helium, deuterium, and BAO data with an almost model-independent prior on $\theta_s$ derived from Planck data. Another recent study on BBN \cite{Peimbert:2016bdg} provide a tight bound of $N_{\textrm{eff}} = 2.90\pm0.22$ (68\% C.L.), which means at at 95\% C.L., there will be only a small overlap in the values of $N_{\textrm{eff}}$ provided by \cite{Peimbert:2016bdg} and TT+lowP+HST. It is also to be noted that addition of the HST prior leads to a slightly inferior fit to the data, due to the $3.4\sigma$ tension present between Planck and HST regarding the value of $H_0$ \cite{0004-637X-826-1-56}. We find that in this $\Lambda \textrm{CDM}+r_{0.05}+N_{\textrm{eff}}+ m^{\textrm{eff}}_{\textrm{s}}$ model, compared to TT+lowP, the dataset TT+lowP+HST degrades the $\chi^2$-fit by an amount of $\Delta \chi^2 = +3.43$. 

\begin{figure}[tbp]
\centering % \begin{center}/\end{center} takes some additional vertical space
\includegraphics[width=.4963\linewidth]{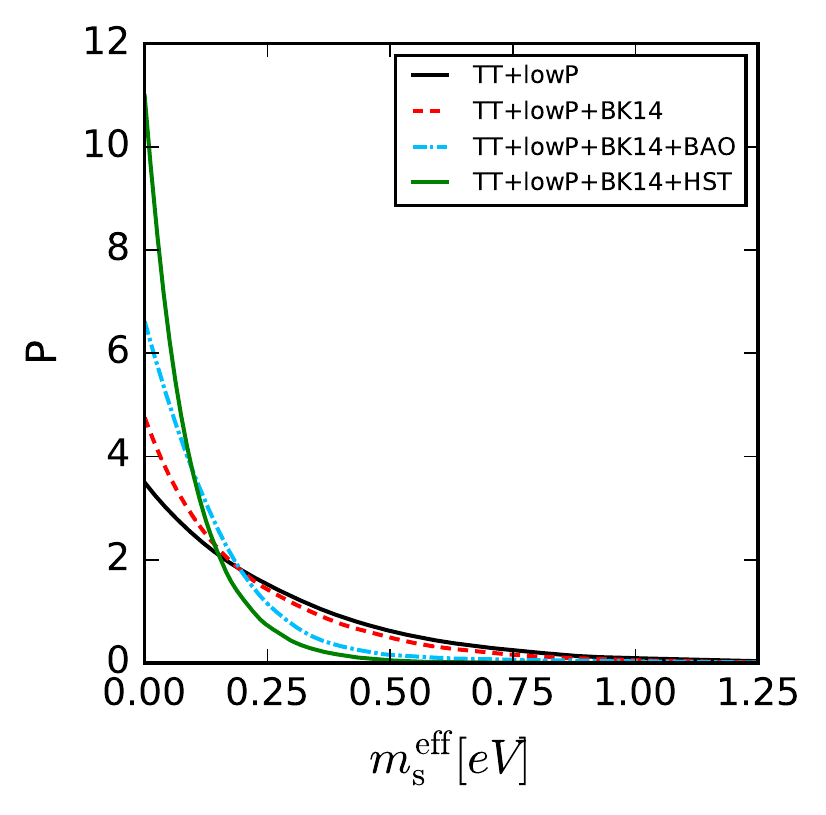}
\hfill
\includegraphics[width=.4963\linewidth]{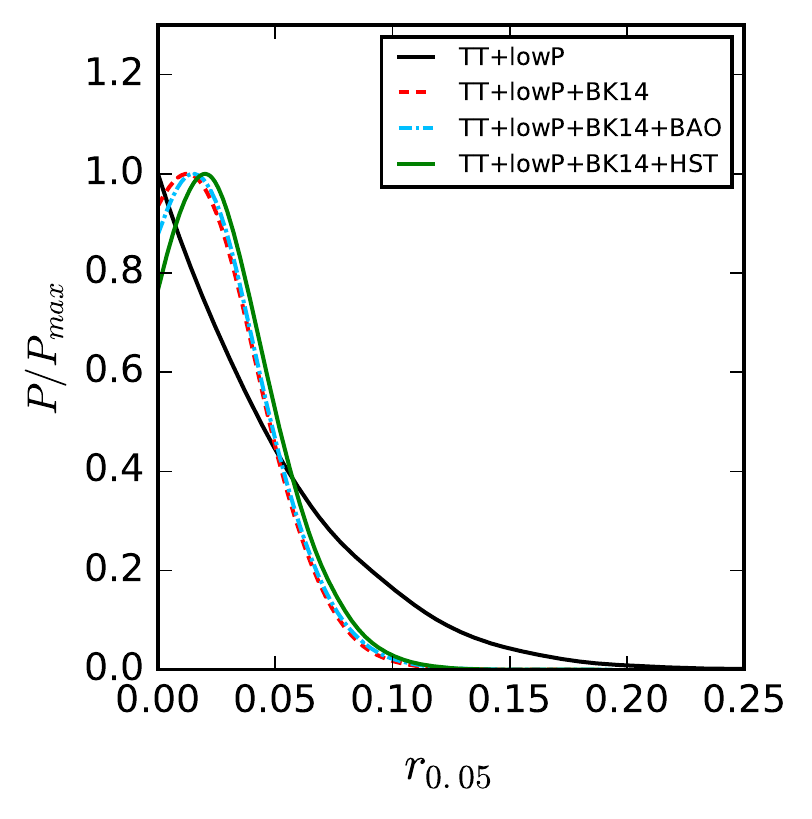}

% "\includegraphics" is very powerful; the graphicx package is already loaded
\caption{\label{fig:2}1-D marginalized posteriors for $m^{\textrm{eff}}_{\textrm{s}}$ [eV] and $r_{0.05}$ in the $\Lambda \textrm{CDM} + r_{0.05} + N_{\textrm{eff}} + m^{\textrm{eff}}_{\textrm{s}}$ model with various data combinations.}
\end{figure}

\emph{Akaike information criterion (AIC)}:
To understand the improvement/worsening of the quality of fit with addition of sterile neutrino parameters ($N_{\textrm{eff}}$ and $m_{\textrm{s}}^{\textrm{eff}}$) we need to compare the fit to data given by $\Lambda \textrm{CDM}+r_{0.05}+N_{\textrm{eff}}+m_{\textrm{s}}^{\textrm{eff}}$ with that of $\Lambda \textrm{CDM} + r_{0.05}$. Since the number of parameters in the two models are not same, a popular method to compare the fit is the Akaike information criterion (AIC) \cite{1100705}.  For a particular model and data, AIC is defined as, 
\begin{equation}
\textrm{AIC} = \chi^2_{\textrm{best-fit}} + 2k
\end{equation}
 where $k$ is the number of parameters in the model. The model with lower AIC corresponds to the preferred model.\\
Thus, comparison with another model (with the same data) can be done with $\Delta \textrm{AIC} = \Delta \chi^2 + 2\Delta k$.
Usually models with extra parameters provide better fit to the data since they have a larger parameter space. The $2\Delta k$ term penalises models with extra parameters to prevent any over-fitting. Here $2\Delta k = 4$.
 
We find that for the TT+lowP+HST data:  
\begin{equation}
\Delta \chi^2 = \chi^2_{\textrm{best-fit}} (\Lambda \textrm{CDM}+r_{0.05}+N_{\textrm{eff}}+m_{\textrm{s}}^{\textrm{eff}}) - \chi^2_{\textrm{best-fit}} (\Lambda \textrm{CDM}+r_{0.05}) = -4.3
\end{equation}
i.e., the $\Lambda \textrm{CDM}+r_{0.05}+N_{\textrm{eff}}+m_{\textrm{s}}^{\textrm{eff}}$ model provides a better $\chi^2$ fit compared to $\Lambda \textrm{CDM} + r_{0.05}$. But due to the 2 extra parameters, $\Delta \textrm{AIC} = -0.3$. Since this difference is small, it implies that the goodness of fits to the TT+lowP+HST data for the two models are similar.\\

\begin{figure}[tbp]
\centering % \begin{center}/\end{center} takes some additional vertical space
\includegraphics[width=.6\linewidth]{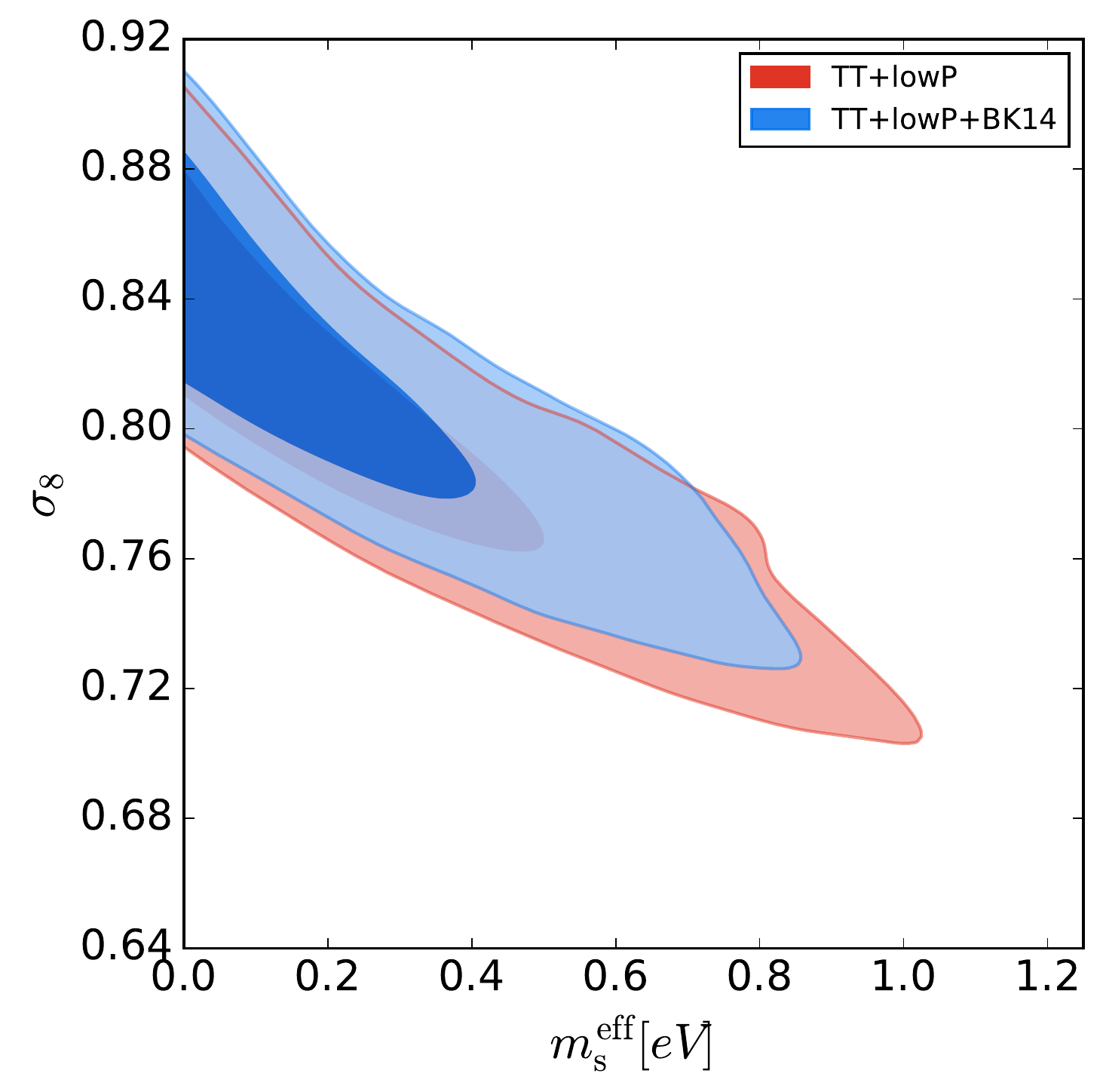}
% "\includegraphics" is very powerful; the graphicx package is already loaded
\caption{\label{fig:3}1$\sigma$ and 2$\sigma$ marginalized  contours for $\sigma_8$ vs. $m^{\textrm{eff}}_{\textrm{s}}$ [eV] in the $\Lambda \textrm{CDM} + r_{0.05} + N_{\textrm{eff}} + m^{\textrm{eff}}_{\textrm{s}}$ model with the following combinations: TT+lowP and TT+lowP+BK14. Adding BK14 leads to slightly higher $\sigma_8$; and due to large anti-correlation present between $\sigma_8$ and $m^{\textrm{eff}}_{\textrm{s}}$, slightly stronger bound on $m^{\textrm{eff}}_{\textrm{s}}$ is obtained.}
\end{figure}

Since the main aim of this paper is to analyze the role of the BK14 data, Table~\ref{table:3} lists the bounds on the cosmological parameters, now with BK14 data included in each combination. The inclusion of the BK14 data seems to have almost no effect on the bounds of $N_{\textrm{eff}}$ and $H_0$, as can be seen by comparing the results of Table~\ref{table:2} and Table~\ref{table:3}. However, bounds on $ m^{\textrm{eff}}_{\textrm{s}}$ improve slightly across all data combinations. The 1-D marginalized posteriors for $m^{\textrm{eff}}_{\textrm{s}}$ and $r_{0.05}$ for various datasets are shown in Figure~\ref{fig:2}. While for TT+lowP, we had $m^{\textrm{eff}}_{\textrm{s}} <$ 0.78 eV, this bound improves to $m^{\textrm{eff}}_{\textrm{s}} <$ 0.68 eV with TT+lowP+BK14. Addition of BAO data further improves this bound to $m^{\textrm{eff}}_{\textrm{s}} <$ 0.46 eV. Our most aggressive bound in this paper comes with TT+lowP+BK14+HST: $m^{\textrm{eff}}_{\textrm{s}} <$ 0.28 eV.
BK14 data significantly constrains the tensor-to-scalar ratio, $r_{0.05}$. TT+lowP provides $r_{0.05}<0.127$ whereas TT+lowP+BK14 gives a constraint of $r_{0.05}<0.068$. However, we found only a very small correlation between $r_{0.05}$ and $m^{\textrm{eff}}_{\textrm{s}}$, and that does not explain the decrease in mass. In fact the correlation coefficient (defined as $R_{ij} \equiv C_{ij}/\sqrt{C_{ii} C_{jj}}$, where $i$ and $j$ are the two parameters being considered and $C$ is the covariance matrix of cosmological parameters) between $r_{0.05}$ and $m^{\textrm{eff}}_{\textrm{s}}$ to be $R_{m^{\textrm{eff}}_{\textrm{s}},r_{0.05}} = -0.08$ with TT+lowP and $R_{m^{\textrm{eff}}_{\textrm{s}},r_{0.05}} = +0.02$ with TT+lowP+BK14, i.e., there is no significant correlation before addition of BK14 and also no significant change after. However we also find slightly increased values of $\sigma_8$ across all data combinations when BK14 is included. For instance, for TT+lowP, we have $\sigma_8 = 0.802^{+0.040}_{-0.029}$, which increases to $\sigma_8 = 0.814^{+0.036}_{-0.027}$ with TT+lowP+BK14. Since $\sigma_8$ is the normalization of matter power spectrum on scales of $8h^{-1}$ Mpc, a higher $\sigma_8$ prefers lower sterile neutrino mass, as larger neutrino masses create larger suppressions in the matter power spectrum \cite{Lesgourgues:2014zoa}. Thus $\sigma_8$ and $m^{\textrm{eff}}_{\textrm{s}}$, both are strongly anti-correlated. Indeed, we found $R_{\sigma_8,m^{\textrm{eff}}_{\textrm{s}}} = -0.84$ with TT+lowP and $R_{r_{0.05},m^{\textrm{eff}}_{\textrm{s}}} = -0.81$ with TT+lowP+BK14, and hence, even such small changes in $\sigma_8$ should also create small changes in $m^{\textrm{eff}}_{\textrm{s}}$, which we find is the case here. This has been depicted in Figure~\ref{fig:3}. Again, notice that the lensing data prefers a lower $\sigma_8$ value. As in Table~\ref{table:3}, TT+lowP+BK14+HST+BAO yields $\sigma_8 = 0.841^{+0.025}_{-0.021}$, whereas adding the lensing data to this combination yields a lower $\sigma_8 =0.820^{+0.019}_{-0.015}$. Due to the same anti-correlation between $\sigma_8$ and $m^{\textrm{eff}}_{\textrm{s}}$, we see that inclusion of lensing data degrades the $m^{\textrm{eff}}_{\textrm{s}}$ bounds.

Overall, we can say that the BK14 data makes the case for fully thermalized eV scale sterile neutrinos slightly worse. The parameter to justify this statement is $m_{\textrm{s}}^{\textrm{eff}}$. As we have shown that addition of the BK14 data does not affect the $N_{\textrm{eff}}$ bounds, BK14 data does not affect the thermalisation situation, as far as cosmological data is concerned. However, short baseline oscillation experiments predict a fully thermalised sterile neutrino of mass $\simeq$ 1 eV. This requires that both $N_{\textrm{eff}} = 4.046$ and $m^{\textrm{eff}}_{\textrm{s}} \simeq 1$ eV be allowed by the data. Since adding the BK14 data tightens the bounds on $m^{\textrm{eff}}_{\textrm{s}}$ for all of the cosmological dataset combinations, it also takes the $m^{\textrm{eff}}_{\textrm{s}}$ value further away from the 1 eV value, while $N_{\textrm{eff}}$ bounds almost remain unchanged. Effect of BK14 data on sum of active neutrino masses ($\sum m_{\nu}$) was also studied by us recently in \cite{Choudhury:2018byy}, in the $\Lambda \textrm{CDM} + r_{0.05} + \sum m_{\nu}$ model, where we had also found slightly increased $\sigma_8$. This is also indirectly confirmed by the recent Planck 2018 results, where they provide a bound of $\sum m_{\nu} <$ 0.12 eV with Planck TT,TE,EE+lowE+lensing+BAO data in $\Lambda \textrm{CDM}+\sum m_{\nu}$ model \cite{Aghanim:2018eyx}, whereas the bound is $\sum m_{\nu} <$ 0.11 eV with Planck TT,TE,EE+lowE+lensing+BK14+BAO data in the $\Lambda \textrm{CDM}+r+\sum m_{\nu}$ model \cite{Akrami:2018odb}. This similar effect was seen to persist even in a 12 parameter extended scenario in a recent study with non-phantom dynamical dark energy \cite{Choudhury:2018adz}. In this paper we have shown that such an effect is also present in an extended $\Lambda \textrm{CDM}$ cosmology with light sterile neutrinos. CMB B-mode polarization has two known sources \cite{Ade:2014xna}. The first one is the inflationary gravitational waves (IGW), i.e., tensors (expected to produce a bump peaked around $l\simeq 80$, the so called 'recombination bump' in the BB-mode CMB spectra) as tensors induce quadruple anisotropies in the CMB within the last scattering surface. The tensor signature cannot be reproduced by scalar perturbations, and the amplitude of the recombination bump depends on the tensor-to-scalar ratio. The second source is gravitational lensing by large scale structure. It leads to deflection of CMB photons at late times, which converts a small part of the E mode power into B mode. This lensing BB spectra is expected to have a peak around $l\simeq 1000$. The BICEP2/Keck experiment has a multipole range $20<l<330$ aiming to constrain the tensor-to-scalar ratio. However since $r_{0.05}$ and $m^{\textrm{eff}}_{\textrm{s}}$ are only weakly correlated, the slightly stronger constraints on the neutrino masses is possibly coming from gravitational lensing information encoded in the BK14 data, and not from measurement of $r_{0.05}$.

\begin{figure}[tbp]
\centering % \begin{center}/\end{center} takes some additional vertical space
\includegraphics[width=.4963\linewidth]{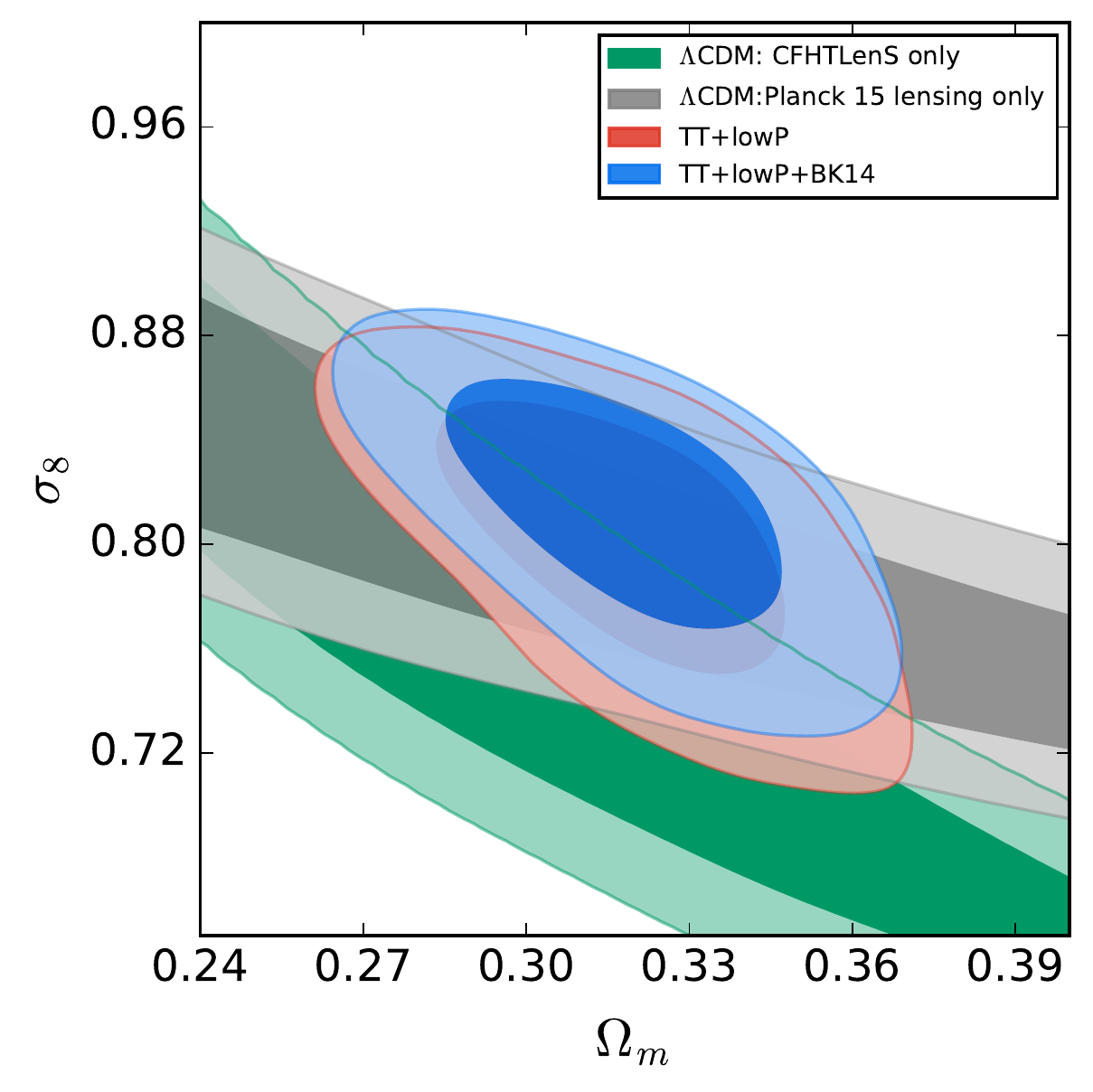}
\hfill
\includegraphics[width=.4963\linewidth]{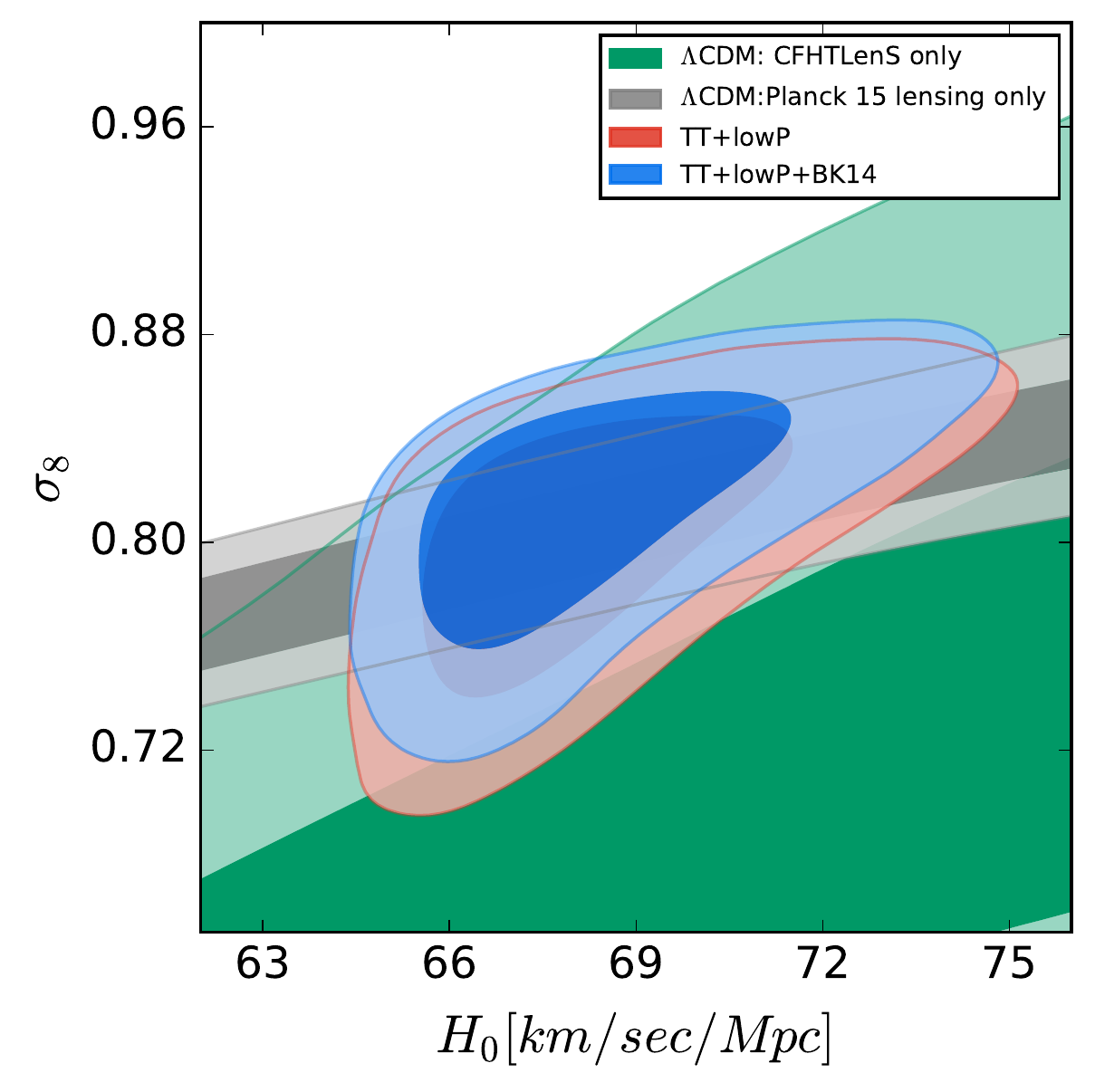}% "\includegraphics" is very powerful; the graphicx package is already loaded
\caption{\label{fig:4}1$\sigma$ and 2$\sigma$ marginalized  contours for $\sigma_8$ vs. $\Omega_m$ and $\sigma_8$ vs. $H_0$ in the $\Lambda \textrm{CDM} + r_{0.05} + N_{\textrm{eff}} + m^{\textrm{eff}}_{\textrm{s}}$ model with the following combinations: TT+lowP and TT+lowP+BK14. We have also presented the contours in the $\Lambda \textrm{CDM}$ model with Planck 2015 lensing and CFHTLenS data. Adding BK14 leads to slightly higher $\sigma_8$, which worsens the agreement with CFHTLenS and Planck 2015.}
\end{figure}

\emph{$H_0$ and $\sigma_8$ tensions:}

It is also worth noting that in $\Lambda \textrm{CDM}$ model, with TT+lowP, Planck collaboration \cite{Planck2015} found that $H_0 = 67.31\pm 0.96$ km/sec/Mpc, whereas in this $\Lambda \textrm{CDM} + r_{0.05} + N_{\textrm{eff}} + m^{\textrm{eff}}_{\textrm{s}}$ model we find $H_0 = 68.35^{+1.23}_{-2.50}$ km/sec/Mpc. This preference to larger values of $H_0$ decreases the more than 3$\sigma$ tension present in the $\Lambda \textrm{CDM}$ model, between Planck 2015 and HST. One of the main reasons is that marginalizing over $N_{\textrm{eff}}$, which allows for $N_{\textrm{eff}}>3.046$ and higher $N_{\textrm{eff}}$ values prefer a higher $H_0$, to keep the acoustic scale parameter $\theta_s$ fixed \cite{Planck2015}, which is very well constrained by Planck data. Thus $H_0$ and $N_{\textrm{eff}}$ are strongly correlated.

The $\Lambda \textrm{CDM} + r_{0.05} + N_{\textrm{eff}} + m^{\textrm{eff}}_{\textrm{s}}$ model also helps in reconciling the $\sigma_8$ tension present in the $\sigma_8-\Omega_m$ plane in $\Lambda \textrm{CDM}$ model between Planck 2015 and weak lensing survey, like CFHTLenS \cite{Heymans:2012gg} and KiDS-450 \cite{Hildebrandt:2016iqg}. For instance, the KiDS-450 survey constrains the quantity $S_8 \equiv \sigma_8 \sqrt{\Omega_m/0.3} = 0.745 \pm 0.039$ which has a 2.3$\sigma$ tension with Planck TT+lowP, which prefers a much higher value of $S_8 = 0.851\pm0.024$ \cite{Planck2015}. Planck data also prefers higher values of $\sigma_8$ compared to CFHTLenS. With TT+lowP in base $\Lambda \textrm{CDM}$ model, one gets $\sigma_8 = 0.829\pm0.014$ \cite{Planck2015}. However, in this $\Lambda \textrm{CDM} + r_{0.05} + N_{\textrm{eff}} + m^{\textrm{eff}}_{\textrm{s}}$ model, with TT+lowP, we get $\sigma_8 = 0.802^{+0.040}_{-0.029}$, which is much lower and thereby the conflict is decreased somewhat. We also get $S_8 = 0.824^{+0.030}_{-0.027}$, which is better agreement with KiDS-450 than $\Lambda \textrm{CDM}$. However, the BK14 data prefers slightly higher $\sigma_8$ values and thereby increases the tension between Planck and these weak gravitational lensing surveys. This can be visualized in Figure~\ref{fig:4}, where we see that the inclusion of BK14 data drives the 2-D contours upwards to a small extent. In Figure~\ref{fig:4}, we have used the CFHTLenS data with conservative cuts as described in \cite{Planck2015}.

Another important point is that while $\Lambda \textrm{CDM} + r_{0.05} + N_{\textrm{eff}} + m^{\textrm{eff}}_{\textrm{s}}$ helps in relieving the $H_0$ and $\sigma_8$ tensions present in the $\Lambda \textrm{CDM}$ model, they are not both relieved together in any region of the allowed parameter space. In the right panel of Figure~\ref{fig:4}, we can see that the regions where $\sigma_8$ has lower values, $H_0$ also has lower values (while we need higher values of $H_0$ to relieve the $H_0$ tension), and similarly, where $H_0$ has higher values, $\sigma_8$ also has higher values (while we need lower values of $\sigma_8$ to relieve the $\sigma_8$ tension). This in turn implies that the two conflicts are not resolved together in this model. And BK14 data worsens the conflicts even more. The HST prior also doesn't help the issue here. As we can see from Tables~\ref{table:2} and \ref{table:3}, the inclusion of this Gaussian prior leads to a preference for much higher $N_{\textrm{eff}}$ values, and higher $\sigma_8$ values as well, increasing the conflict.

\subsection{Results for $\Lambda \textrm{CDM} + r_{0.05} + m^{\textrm{eff}}_{\textrm{s}}$ model}
\label{sec:3:2}
\begin{table}
	
	\centering
%	\small\addtolength{\tabcolsep}{-4pt}
%   \resizebox{\textwidth}{!}{
	\begin{tabular}{cccc}
		\toprule
		\toprule
		 %\vspace{0.2cm}
		Parameter &    TT+lowP  &    TT+lowP+BK14\\ \midrule  \vspace{0.2cm}
		%\vspace{0.2cm}
		$m^{\textrm{eff}}_{\textrm{s}}$ (eV)  &    $<0.66$  &    $<0.50$\\ \vspace{0.2cm}
		
	 	$r_{0.05}$   &    $<0.175$  &    $<0.076$\\ \vspace{ 0.2cm}	
	
	 	$H_0$ (km/sec/Mpc)  &    $73.92^{+2.60}_{-1.37}$  &    $74.20^{+2.13}_{-1.28}$\\ \vspace{0.2cm}
		
	 	$\sigma_8$  &    $0.840^{+0.049}_{-0.020}$  &    $0.857^{+0.039}_{-0.018}$\\ 
		\bottomrule
		\bottomrule
	\end{tabular}
	\caption{\label{table:4} \footnotesize Bounds on a cosmological parameters in the $\Lambda \textrm{CDM} + r_{0.05} + m^{\textrm{eff}}_{\textrm{s}}$ model with $N_{\textrm{eff}}=4.046$, assuming complete thermalization of sterile neutrinos. Marginalized limits are given at 68\% C.L. whereas upper limits are given at 95\% C.L. Note that $H_0$ and $\sigma_8$ are derived parameters.}
\end{table}	
\begin{table}
	
	\centering
%	\small\addtolength{\tabcolsep}{-4pt}
%   \resizebox{\textwidth}{!}{
	\begin{tabular}{ccc}
		\toprule
		\toprule
		 %\vspace{0.2cm}
		Parameter &    TT+lowP  &    TT+lowP+BK14\\ \midrule  \vspace{0.2cm}
		%\vspace{0.2cm}
		$m^{\textrm{eff}}_{\textrm{s}}$ (eV)  &    $<0.83$  &    $<0.63$\\ \vspace{0.2cm}
		
	 	$r_{0.05}$   &    $<0.136$  &    $<0.070$\\ \vspace{ 0.2cm}	
	
	 	$H_0$ (km/sec/Mpc)  &    $69.04^{+2.15}_{-1.59}$  &    $69.25^{+1.94}_{-1.42}$\\ \vspace{0.2cm}
		
	 	$\sigma_8$  &    $0.803^{+0.051}_{-0.025}$  &    $0.820^{+0.041}_{-0.021}$\\ 
		\bottomrule
		\bottomrule
	\end{tabular}
	\caption{\label{table:5} \footnotesize Bounds on a cosmological parameters in the $\Lambda \textrm{CDM} + r_{0.05} + m^{\textrm{eff}}_{\textrm{s}}$ model with $N_{\textrm{eff}}=3.5$, assuming partial thermalization of sterile neutrinos. Marginalized limits are given at 68\% C.L. whereas upper limits are given at 95\% C.L. Note that $H_0$ and $\sigma_8$ are derived parameters.}
\end{table}	

In this section we verify the stability of the results obtained in the previous section, by going to a smaller parameter space. We stop varying $N_{\textrm{eff}}$ and fix its value to 4.046 and 3.5. The first one corresponds to complete thermalization of sterile neutrinos, while the later one corresponds to partial thermalization. We have restricted ourselves to CMB data only. For $N_{\textrm{eff}} = 4.046$ and $N_{\textrm{eff}} = 3.5$, the results are given in Tables~\ref{table:4} and \ref{table:5} respectively.

We see that BK14 does help in obtaining better constraint on the sterile mass also in this reduced parameter space. For $N_{\textrm{eff}} = 4.046$, with TT+lowP, we get $m^{\textrm{eff}}_{\textrm{s}} < 0.66$ eV, whereas inclusion of BK14 leads to a tighter bound of $m^{\textrm{eff}}_{\textrm{s}} < 0.50$ eV. Similar case of strengthening of mass bound is seen with $N_{\textrm{eff}} = 3.5$, although these bounds are more relaxed compared to the case $N_{\textrm{eff}} = 4.046$, as a higher $N_{\textrm{eff}}$ prefers a higher $H_0$. Again we see that the BK14 data itself does not affect the $H_0$ constraints much, but heavily constraints the tensor-to-scalar ratio, and also slightly increases the preferred $\sigma_8$ values. The main conclusions made in the previous section on the larger parameter space thus remains unchanged in this smaller parameter space. 

It is imperative to note that for sterile neutrinos produced by a thermal process and obeying Eq.~\ref{eqn:3}, for $N_{\textrm{eff}} = 4.046$, we have $m^{ph}_s = m^{\textrm{eff}}_{\textrm{s}}$, whereas for $N_{\textrm{eff}} = 3.5$, we have $m^{ph}_s = 1.8 m^{\textrm{eff}}_{\textrm{s}}$. Hence, for $N_{\textrm{eff}} = 3.5$ and with TT+lowP+BK14, we have a corresponding bound of $m^{ph}_s<1.13$ eV. This implies that CMB data allows sterile neutrinos with mass $\simeq 1$ eV, but only with partial thermalization with $N_{\textrm{eff}} \simeq 3.5$. 
When we compare the quality of fit to the TT+lowP+BK14 data between the $\Lambda \textrm{CDM}+r_{0.05}+m_{\textrm{s}}^{\textrm{eff}}$ model ($N_{\textrm{eff}} = 3.5$ and $4.046$) and the $\Lambda \textrm{CDM}+r_{0.05}$ (with $N_{\textrm{eff}} = N_{\textrm{eff}}^{\textrm{SM}}$), we find that, \\
for the $N_{\textrm{eff}} = 4.046$ case:
\begin{equation}
\Delta \chi^2 = \chi^2_{\textrm{best-fit}} (\Lambda \textrm{CDM}+r_{0.05}+m_{\textrm{s}}^{\textrm{eff}}) - \chi^2_{\textrm{best-fit}} (\Lambda \textrm{CDM}+r_{0.05}) = +7.03
\end{equation}
whereas, for the  $N_{\textrm{eff}} = 3.5$ case:
\begin{equation}
\Delta \chi^2 = \chi^2_{\textrm{best-fit}} (\Lambda \textrm{CDM}+r_{0.05}+m_{\textrm{s}}^{\textrm{eff}}) - \chi^2_{\textrm{best-fit}} (\Lambda \textrm{CDM}+r_{0.05}) = -0.22
\end{equation}
 
These correspond to $\Delta \textrm{AIC} = +9.03$ (for $N_{\textrm{eff}} = 4.046$) and $\Delta \textrm{AIC} = +1.78$ (for $N_{\textrm{eff}} = 3.5$). Thus, the model with partial thermalization of $N_{\textrm{eff}} = 3.5$ provides only a slightly worse fit to the data compared to the $\Lambda \textrm{CDM}+r_{0.05}$ model (with $N_{\textrm{eff}} = N_{\textrm{eff}}^{\textrm{SM}}$), and is preferred by the data much more than the full-thermalization case. This is not surprising as in the previous section we had seen that CMB data alone did not allow complete thermalization. 

\section{Discussion}
\label{sec:4}

Short Baseline (SBL) Oscillation anomalies have hinted towards a fully thermalized sterile neutrino with mass around 1 eV. In this paper we have studied, for the first time, the light eV scale sterile neutrino situation in cosmology in light of the BICEP2/Keck array 2014 CMB B-mode polarization data. We call this dataset BK14. We first considered an extended$-\Lambda \textrm{CDM}$ scenario with tensor perturbations and sterile neutrino parameters: $\Lambda \textrm{CDM} + r_{0.05} + N_{\textrm{eff}} + m^{\textrm{eff}}_{\textrm{s}}$ model. Apart from BK14, we have used Planck 2015 temperature and low-$l$ polarization data (TT+lowP), latest BAO data and a Gaussian prior on the Hubble constant (HST) from local measurements. We find that inclusion of the BK14 data has almost no effect on the bounds of $N_{\textrm{eff}}$ and $H_0$ but it strengthens the bounds on $m^{\textrm{eff}}_{\textrm{s}}$ to a small extent by preferring slightly higher values of $\sigma_8$, with which $m^{\textrm{eff}}_{\textrm{s}}$ is strongly anti-correlated. The BK14 data also tightly constraints the tensor-to-scalar ratio, $r_{0.05}$ but we find negligible correlation between $r_{0.05}$ and $m^{\textrm{eff}}_{\textrm{s}}$. This makes us think that the effect on mass is coming from the gravitational lensing information encoded in the B-mode polarization and not from the Inflationary Gravitational Waves. The bound of $m^{\textrm{eff}}_{\textrm{s}} <$ 0.46 eV (95\% C.L.) is found for the combination of Planck 2015, BAO and BK14 datasets, whereas the bound is $m^{\textrm{eff}}_{\textrm{s}} <$ 0.53 eV (95\% C.L.) without the BK14 data. Our most aggressive bound of $m^{\textrm{eff}}_{\textrm{s}} <$ 0.28 eV (95\% C.L.) is obtained with Planck 2015, HST and BK14. The HST prior also leads to high $N_{\textrm{eff}}$ values which allow full thermalization of the sterile neutrino (at 2$\sigma$) but such high values are in conflict with bounds from Big Bang Nucleosynthesis. Also, addition of the HST prior to the TT+lowP data leads to a slightly worse $\chi^2$ fit to the data. On the other hand, it is to be noted that as per the Akaike information criterion (AIC) the $\Lambda \textrm{CDM}+r_{0.05}+N_{\textrm{eff}}+m_{\textrm{s}}^{\textrm{eff}}$ model provides equally good fit to the data as the $\Lambda \textrm{CDM}+r_{0.05}$ model, for the TT+lowP+HST data combination. Previous studies have indicated that fully thermalized sterile neutrinos with mass $\sim 1$ eV (as predicted by SBL experiments) are disfavoured by cosmological data. Our analysis indicates that it becomes slightly more disfavoured with the inclusion of BK14 data, due to tighter mass bounds. The BK14 data also seems to make the agreement between Planck 2015 and CFHTLenS (weak gravitational lensing data) worse due to the higher $\sigma_8$ values. 

We would also like to mention that the Planck 2018 results, released during the preparation of this article, indirectly show tightening of bounds on $\sum m_{\nu}$   with BK14. They provide a bound of $\sum m_{\nu} <$ 0.12 eV with Planck TT,TE,EE+lowE+lensing+BAO data in $\Lambda \textrm{CDM}+\sum m_{\nu}$ model \cite{Aghanim:2018eyx}, whereas the bound is $\sum m_{\nu} <$ 0.11 eV with Planck TT,TE,EE + lowE + lensing + BK14 + BAO data in the $\Lambda \textrm{CDM}+r+\sum m_{\nu}$ model \cite{Akrami:2018odb}. Thus we expect our main conclusion regarding BK14 helping in improving the bound on sterile neutrino mass will remain unchanged if used with the recent Planck 2018 data instead of Planck 2015 that we have used in this paper. 

While this work was still being completed, a new B-mode polarisation data was released publicly, from the same BICEP2/Keck collaboration. This newly released data includes all the measurements upto and including 2015, and thus we call it BK15 \cite{Ade:2018gkx}. To understand the effect of the new data, we performed an MCMC analysis with TT+lowP+HST+BK15 in the $\Lambda \textrm{CDM}+r_{0.05}+N_{\textrm{eff}}+ m^{\textrm{eff}}_{\textrm{s}}$ model (with all other settings remaining unchanged). We found the following bounds: $m^{\textrm{eff}}_{\textrm{s}}<0.27$ eV (95\% C.L.), $r_{0.05} < 0.061$, and $\sigma_8 = 0.847^{+0.026}_{-0.021}$. In the same model, when we had used BK14 instead of BK15, we had found (see table 3), $m^{\textrm{eff}}_{\textrm{s}}<0.28$ eV (95\% C.L.), $r_{0.05} < 0.073$, and $\sigma_8 = 0.846^{+0.026}_{-0.020}$. As we can see, that while the bound on $r_{0.05}$ changes, the bounds on $m^{\textrm{eff}}_{\textrm{s}}$ and $\sigma_8$ almost remain unchanged. We also checked that other parameters of interest, like $H_0$ and $N_{\textrm{eff}}$ change negligibly. As before, since $r_{0.05}$ and $m^{\textrm{eff}}_{\textrm{s}}$ have only a very weak correlation, it doesn't affect the mass bound. On the other hand, since TT+lowP+HST+BK15 almost doesn't change the bound on $\sigma_8$, the mass bound almost remains the same. Thus, we find that reanalysis with BK15 instead of BK14 will not change the neutrino mass bounds.

This tension between SBL and cosmological datasets has given rise to a number ideas to reconcile the eV-scale sterile neutrinos with cosmology. These include introduction of new "secret interactions" among sterile neutrinos which modifies the background potential and blocks thermalization \cite{Archidiacono:2014nda,Forastieri:2017oma,Song:2018zyl,Bento:2001xi,Dasgupta:2013zpn,Hannestad:2013ana,Archidiacono:2016kkh,Chu:2015ipa,Saviano:2014esa,Jeong:2018yts}, modifications to the cosmic expansion rate at the time where sterile neutrinos are produced \cite{Rehagen:2014vna}, large lepton asymmetry \cite{Foot:1995bm,Chu:2006ua,Saviano:2013ktj}, time varying dark energy component \cite{Giusarma:2011zq}, very low reheating temperature \cite{Gelmini:2004ah}. The recent results that have come from the MiniBooNE collaboration \cite{Aguilar-Arevalo:2018gpe} have rekindled interest in the sterile neutrinos. Thus, there seems to be interesting and exciting times ahead in the realm of light sterile neutrinos in cosmology.

\acknowledgments
The authors sincerely thank the anonymous referee for the thoughtful comments and efforts towards
improving our manuscript. SRC thanks the cluster computing facility at HRI (http://cluster.hri.res.in). SRC also thanks Steen Hannestad for useful discussions. The authors would also like to thank the Department of Atomic Energy (DAE) Neutrino Project of HRI. This project has received funding from the European Union's Horizon 2020 research and innovation programme InvisiblesPlus RISE under the Marie Sklodowska-Curie grant agreement No 690575. This project has received funding from the European Union's Horizon 2020 research and innovation programme Elusives ITN under the Marie Sklodowska-Curie grant agreement No 674896.

\bibliography{paper}
\bibliographystyle{jhep}

\end{document}